\documentclass{aa}

\usepackage{psfig}
\usepackage{graphicx}
\usepackage{natbib}
\usepackage{latexsym}
\usepackage{amssymb}
\bibpunct{(}{)}{;}{a}{}{,}

\newcommand{\kms}{km\,s$^{-1}$}

\newcommand{\fefiv}{\mbox{Fe\,5270}}
\newcommand{\fesix}{\mbox{Fe\,5335}}
\newcommand{\hb}{\mbox{H$\beta$}}
\newcommand{\hd}{\mbox{H$\delta$}}
\newcommand{\hdf}{\mbox{H$\delta_F$}}
\newcommand{\hda}{\mbox{H$\delta_A$}}
\newcommand{\hg}{\mbox{H$\gamma$}}
\newcommand{\hgf}{\mbox{H$\gamma_F$}}
\newcommand{\hga}{\mbox{H$\gamma_A$}}

\newcommand{\mgb}{\mbox{Mg$_b$}}
\newcommand{\oiii}{O\,\textsc{iii}}

\begin{document}

\title{
Kinematic and chemical evolution of early--type galaxies
\thanks{Based on observations with the European Southern
Observatory Very Large Telescope (ESO-VLT), 
observing run IDs 65.O-0049, 66.A-0547 and 68.A-0013.} 
\thanks{Based in parts on observations (PID\,9502) with the NASA/ESA
  {\it Hubble Space Telescope}, obtained at the Space Telescope
  Science Institute, which is operated by AURA, Inc., under
  NASA contract NAS\,5-26555.}
}

\author{
B. L. Ziegler\inst{1} \and 
D. Thomas\inst{2} \and 
A. B\"ohm\inst{1} \and 
R. Bender\inst{2}\inst{,}\inst{3} \and 
A. Fritz\inst{1} \and 
C. Maraston\inst{2}
}

\offprints{B. L. Ziegler, \\ e-mail: bziegler@uni-sw.gwdg.de}

\institute{Universit\"atssternwarte G\"ottingen, Geismarlandstra{\ss}e 11,
           37083 G\"ottingen, Germany
\and Max--Planck--Institut f\"ur extraterrestrische Physik, 
           Giessenbachstra{\ss}e, 85748 Garching, Germany
\and Universit\"atssternwarte M\"unchen, Scheinerstra{\ss}e 1,
            81679 M\"unchen, Germany
}

\date{Received 02/12/2004 / Accepted 07/12/2004}

\abstract{ We investigate in detail 13 early--type field galaxies with
$0.2<z<0.7$ drawn from the F{\sc ors} Deep Field.  Since the majority
(9 galaxies) is at $z\approx0.4$, we compare the field galaxies to 22
members of three rich clusters with $z=0.37$ to explore possible
variations caused by environmental effects.  We exploit VLT/FORS
spectra ($R\approx1200$) and HST/ACS imaging to determine internal
kinematics, structures and stellar population parameters.  From the
Faber--Jackson and Fundamental Plane scaling relations we deduce a
modest luminosity evolution in the $B$--band of 0.3--0.5\,mag for both
samples.  
We compare measured Lick absorption line
strengths (\hd, \hg, \hb, \mgb, \& \fesix) with evolutionary stellar
population models to derive light-averaged ages, metallicities and the
element abundance ratios Mg/Fe. We find that all these three stellar
parameters of the distant galaxies obey a scaling with velocity
dispersion (mass) which is very well consistent with the one of local
nearby galaxies.  In particular, the distribution of Mg/Fe ratios of
local galaxies is matched by the distant ones, and their derived mean
offset in age corresponds to the average lookback time.  This
indicates that there was little chemical enrichment and no significant
star formation within the last $\sim5$\,Gyr.  
The calculated luminosity evolution of a simple stellar population model
for the derived galaxy ages and lookback times is in most cases very 
consistent with the mild brightening measured by the scaling relations.
%
%
}

\titlerunning{Kinematic and chemical evolution of early--type galaxies}
\authorrunning{B. L. Ziegler et al.}

\maketitle


\section{\label{intro}Introduction}

Recently, it has become feasible to study physical parameters of
galaxies out to redshifts $z\approx1$.  Spectroscopy at 10m-class
telescopes allows to determine the internal kinematics like rotational
speed and velocity dispersion as well as the age and metallicity
distribution of the stellar populations.  Spatially high-resolution
imaging with the Hubble Space Telescope (HST) reveals morphologies,
characteristic sizes and total luminosities of distant galaxies.  In
particular powerful are scaling relations that combine the kinematics
(sampling both dark and bright matter) and the stellar populations
(baryons).  For dynamically hot galaxies, such scaling relations are
the Faber--Jackson relation \citep[FJR;][]{FJ76} between the velocity
dispersion $\sigma$ and the absolute luminosity and the Fundamental
Plane \citep[FP;][]{DD87,DLBDFTW87} encompassing effective radius
$R_e$, effective surface brightness $\langle\mu\rangle_e$, and
$\sigma$.  The chemical enrichment history of the galaxies can be
examined by comparing absorption line strengths to evolutionary
stellar population models.

With these observational quantitative measurements, it is now possible
to perform robust tests of galaxy formation theories.  The basic
prediction of cold dark matter (CDM) models of hierarchically growing
structure is the gradual increase of mass of galaxies.  Connecting
this mass assembly with the physics of star formation leads to further
predictions for the evolution of the stellar populations of a galaxy.
The combination of characteristic velocities and sizes could in
principle constrain the mass of galaxies, but this transformation is
degenerate.  The main problem is the cutoff radius for the dark matter
halo which must follow certain assumptions by considering various
constraints which greatly affects the estimated total mass of a
galaxy.  Thus, observations can better be taken to investigate the
luminosity and chemical evolution and the age distributions of the
stellar populations.

For spheroidal (elliptical) galaxies, CDM simulations yield a
dependence on environment \citep[e.g.][]{BCF96}.  Galaxies born in
high-density regions that evolve into rich clusters are cutoff from
external influences much earlier on average than those in the field.
Two third of the field ellipticals should still get produced at
$z<1$ \citep[e.g.][]{KCW97}, whereas the last major merger between two
large galaxies in high density environments ending up in a big
elliptical is predicted to occur at somewhat higher redshifts
\citep[e.g.][]{Kauff96,CLBF00}. These predictions manifest themselves
in measureable parameters like mean age, metallicity and element
abundances on the one side and a specific luminosity and color
evolution on the other side.  Thus, field ellipticals, in particular
the bright ones, should on average have younger global ages
\citep[e.g.][]{KC98} and more solar-like element abundance
ratios \citep[e.g.][]{Thoma99} than cluster ellipticals.

The FJR \& FP scaling relations \citep[e.g.][]{BBF92}, 
the color-magnitude \citep[e.g.][]{BLE92} and 
Mg-$\sigma$ relations \citep[e.g.][]{BBF93} 
of \textit{local} early-type galaxies are very tight indicating that the 
bulk of their stars must have been formed at high redshifts ($z\gtrsim2$).
Nevertheless, this homogeneity can be reconciled with the hierarchical
merging scenario \citep{Kauff96}.
Difficulties arise because of a probable degeneracy between age and
metallicity effects, in particular for the Mg-$\sigma$ relation
\citep[e.g.][]{TFWG00a,KSCDK02,TMBM04}.

But the Fundamental Plane is very well suited to study the \textit{redshift 
evolution} of the luminosity \citep[e.g.][]{BSZBBGH97b} and 
the mass/light ratios \citep[e.g.][]{Franx93b}.
Observations up to $z\approx1$ were predominantly carried out for
(bright and massive) galaxies in clusters.
Here, consistent results were obtained of a very modest brightening and
slow decrease of $M/L$ with $z$ fully compatible with the assumption of a very
high formation redshift \citep[e.g.][]{DFKI98,KIDF00b,DS03}.
On the other hand, findings for high-$z$ field galaxies are more discrepant.
Some studies argue for a strong evolution \citep[e.g.][]{TSCMB02,GFKIS03}
whereas other authors favor a behaviour very similar to cluster galaxies
\citep[e.g.][]{DE03,WFDR04}.
The differences might mainly come from the low number of observed
galaxies.
But the samples might also be affected by the so-called progenitor bias 
\citep{DF01} in the sense that the selection criteria for early-type galaxies 
differ strongly among researchers.

Many stellar population studies of the age and metallicity distribution have 
been carried out for \textit{local} galaxies.
The majority of cluster ellipticals have very old mean ages
($\approx10$\,Gyr), high metallicities ($\approx1-3\times Z_\odot$)
and are enhanced in Mg over Fe compared to the solar ratio
([Mg/Fe]\,$\approx0.3$)
\citep[e.g.][]{TFWG00a,PBCMD01,MTSBW03,EHFNB03,TMBM04}.  There are
trends between these parameters and the luminosity with less luminous
galaxies having a wider spread and being skewed to younger ages.  A
more diverse behaviour is found for S0 galaxies in clusters.  At least
two families can be distinguished with one being very similar to the
ellipticals and the other having younger luminosity-weighted ages
($\approx5$\,Gyr) indicating recent star formation activity.  For
field early-type galaxies, \citet{KSCDK02} find that they are younger
(by $2-3$\,Gyr) but more metal-rich (by $\approx0.2$ dex) than cluster
galaxies exhibiting similar super-solar Mg/Fe ratios.  In their recent
study of field and cluster galaxies compiling high-quality data from
several sources, \citet{TMBM04} show that mean ages, metallicities,
and $\alpha$/Fe ratios correlate with galaxy mass ($\sigma$).  Both
zero-point and slope of the $\alpha$/Fe--$\sigma$ relation are
independent of the environmental density, while the ages of objects in
low density environments appear systematically lower accompanied by
slightly higher metallicities.

Little work has been done yet investigating ages and metallicities
from absorption lines in \textit{distant} galaxies because the high
$S/N$ needed for such an analysis is achieved with only long exposures
even at big telescopes and because individual lines can be severely
corrupted by terrestrial spectral features in the red wavelength
regime.  But models of galaxy formation can be constrained much
stronger by measurements at higher redshifts because differences among
the predictions are much larger.  \citet{JSC00}, for example, use
combinations of \hd\ with metal lines measured in co-added spectra of
several E and S0 galaxies in clusters at $z=0, 0.3, 0.5$,
respectively.  They claim that there is no significant difference in
the age--metallicity distribution between these two galaxy types at
any redshift.  \citet{KIFD01} presented a study of the evolution of
the \hg\ and \hd\ Balmer absorption in galaxies in four clusters up to
$z=0.8$.  Assuming that the same relation between these indices and
velocity dispersion holds at all redshifts, they derived a modest
evolution in the zero-point as expected for a passive evolution of old
stellar populations and consistent with the evolution of $M/L$
determined from an FP analysis.  Recently, \citet{EHFNB03} examined
22,000 bright early-type galaxies from the SDSS averaging the spectra
within bins of luminosity, environment, and redshift ($0<z<0.5$) to
produce high-$S/N$ mean spectra.  They confirmed that these $L_*$
galaxies have mainly old quiescent stellar populations with high
metallicities and an excess of $\alpha$- (Mg-) elements with respect
to the solar value independent of environment or redshift.

In this paper, we analyse the properties of galaxies at redshifts
$z\approx0.4$ both in clusters and in the field.  In Section 2 our
observations and data analysis is described.  The Faber--Jackson
relation and the Fundamental Plane are used to investigate the
evolution in luminosity in Section 3.  Measured absorption line
strengths (Section 5) are compared in Section 6 to stellar population
models that explicitely take into account variations in the abundance
of elements to explore the chemical enrichment histories of the
galaxies.  A summary and discussion is presented in Section 7.

Throughout the paper, we adopt the ``concordance'' cosmology with
$\Omega_{\rm matter}=0.3$, $\Omega_{\lambda}=0.7$, and 
$H_0 = 70$\,km\,s$^{-1}$\,Mpc$^{-1}$
\citep[e.g.][]{TSBCC03}. 
This yields a distance luminosity of
104.75 for the Coma cluster which is used as a local reference sample.


\section{\label{obs}Observations and data analysis}

\subsection{Elliptical galaxies from the F{\sc ors} Deep Field}

The spectroscopic observations of our sample of field ellipticals were 
performed in parallel to those of
late--type galaxies in the F{\sc ors} Deep Field that form the basis for
our analysis of the evolution of the Tully--Fisher relation
\citep{ZBFJN02,BZSBF03}.
The main differences are that the early--type galaxy candidates were placed
onto slitlets independent of their position angles and were observed with
more than one MOS setup in many cases to increase the respective exposure
time. 
All in all, nine different setups with the FORS1\&2 instruments at the VLT 
were acquired in three different observing runs with a total integration time
of 2.5 hours for each setup.
Typically, each setup contained 2--4 elliptical candidates.

Object selection was based on the deep $UBgRI$ photometry of the F{\sc ors} 
Deep Field (FDF), which is described e.g. in \citet{HAGJS03}.
Candidates were chosen according to their spectrophotometric type and their
extended structureless appearance. 
Main criterium was the apparent brightness ($R\le22.0$) in order to 
ensure sufficient $S/N$ in the absorption lines for a robust determination
of line strengths and velocity dispersions.
Photometric redshifts were restricted to $z_p<0.6$ of objects with an 
early model SED type (for a description of the photometric redshift 
performance see e.g. \citet{BABDF01}).
Although the twodimensional distribution of FDF objects indicated a cluster of
galaxies with $z_p\approx0.33$ and our primary goal was to target 
\textit{field} ellipticals, we did not preselect against such candidates.
After the spectral analysis, we could confirm that 15 out of
32 observed different elliptical galaxy candidates are most probably members 
of a cluster.
Taking our measurements of the radial velocities of these galaxies, the lower 
limit for the velocity dispersion of the cluster is $\sigma_c\gtrsim430$\,\kms.
Since the cluster center was not observed but only the outer parts, the true
$\sigma_c$ is probably larger.
At $z=0.33$, the Mg\,5170 absorption feature is unfortunately corrupted due
to terrestrial absorption (the B band)
making it impossible to accurately determine the internal velocity dispersion
of a galaxy (see below).
For that reason we do not include these cluster galaxies in our analysis and, 
from hence forward, when we speak of FDF galaxies only the field ellipticals 
are meant.

The elliptical candidates were spread out among the spiral candidates for
each setup placing them on a region of the CCD corresponding to observed
wavelengths where either the \mgb-feature ($\lambda_0\approx5170$\,\AA) or
the G-band ($\lambda_0\approx4300$\,\AA) should have been visible according 
to the photometric redshifts.
As is described in more detail in \citet{BZSBF03} the MOS spectroscopy was
performed using grism \textsf{600R} with order separation filter 
\textsf{GG435} with FORS2 (September and October 2000) and FORS1 (October
2001) at the VLT.
With slit widths of 1\arcsec, a spectral resolution of $R\approx1200$ was
achieved, the spatial scale was 0.2\arcsec/pixel, and seeing conditions
were sufficient (varying between 0.4\arcsec\ and 0.9\arcsec\ FWHM) to meet
the Nyquist theorem.

Image reduction followed the usual steps of bias subtraction, flat fielding,
sky subtraction, and wavelength calibration. Because some spectra exhibited
spatial distortions, the images of each individual slitlet were extracted
from the full frame after bias subtraction allowing the typical 
two-dimensional image reduction of long-slit spectroscopy. All images of
one respective slitlet including the science, flatfield, and arclamp frames
were rectified in exactly the same manner with the procedure described in
\citet{JZBHM04} to ensure correct wavelength calibration.
The spectral rows of an object were averaged using a Horne-based
\citep{Horne86} algorithm as implemented in the image analysis software
MIDAS\footnote{ESO-MIDAS (Munich Image Data Analysis System) is developed and 
maintained by the European Southern Observatory.}
(see e.g. \citet{ZBSDL01}).
Finally, the one-dimensional spectra of the respective exposures were
summed up. 
In those cases where a galaxy was observed in two or three different MOS
setups, the wavelength ranges covered did not match each other exactly
leading to varying final count rates at different wavelengths.

The internal velocity dispersions $\sigma$ and the radial velocities $v$ 
of the galaxies were determined applying the \textit{Fourier 
Correlation Quotient} (FCQ) method with an updated version of \citet{Bende90a}.
For each galaxy, $\sigma$ and $v$ were measured using 3--4 different
kinematic standard stars. 
Deviations from star to star ranged from 2 to 20\,\kms. 
Since only one good template star (SAO\,162947) was available that was 
observed with FORS like the galaxies, 
three more K giant stars (SAO\,32042, SAO\,80333 \& SAO\,98087) were used 
which had been observed with MOSCA at the 3.5m-telescope at the Calar Alto 
observatory\footnote{The German--Spanish Astronomical Center (DSAZ or CAHA), 
operated by the 
Max--Planck--Institut f\"ur Astronomie, Heidelberg, jointly with the Spanish 
National Commission for Astronomy.}
and have sufficiently resolved spectra ($\sigma_*\approx55$\,\kms\ around
\mgb).
FCQ was run on either a ``red'' part of a galaxy spectrum (ideally covering 
\hb\ --- \fesix, but sometimes only \mgb\ or only \hb\ could solely be used) 
or a 
``blue'' part (around the G-band), in some cases (5 out of 13) in both regimes.
In order to determine the instrumental resolution of the galaxy spectra, the
widths of 3--5 emission lines of the respective arc spectrum used for the 
wavelength calibration were measured at wavelengths corresponding to \mgb\ and 
the G-band separately. 
Typical instrumental resolutions are $\sigma_i\approx90-100$\,\kms.
To provide the ``correct'' template for the respective instrumental 
broadening, each stellar 
spectrum was artificially smeared out to $\sigma_i$ before its usage in FCQ.
In the Appendix, we tabulate the error-weighted averages of $\sigma$ and
$v$ (corrected for the radial velocities of the template stars) and their
respective errors for 13 FDF field galaxies that had sufficient $S/N$.
We give only either the values derived from the blue or the red spectral
part depending on their quality, since only these values are used for the
analysis below.
The signal-to-noise of a galaxy's continuum was determined with FCQ,
too, which was calibrated with Monte-Carlo simulations using the stellar 
spectra. 

Absorption line strengths as defined in the Lick system \citep{WO97,TWFBG98}
were measured as described in \citet{ZBSDL01}. 
To this purpose, the galaxy spectra had been artificially broadened first to 
match the instrumental resolution of the stellar Lick/IDS spectra 
($\sigma_{\mathrm{IDS}}\approx210$\,\kms\ around \mgb) before the equivalent
widths were calculated taking the redshifts into account. 
The measured values were then corrected for the decrease in strength caused by
the broadening due to the galaxies' internal velocity dispersions. 
The correction functions for the different absorption lines were determined
by simulating this effect with the stellar spectra.
Lick indices of the galaxies can be badly affected if either the central 
bandpass or those that sample the pseudo-continuum of a line are redshifted 
into the region where telluric emission lines are so strong that the residual 
spectrum after sky subtraction is too noisy. 
A line gets totally corrupted, if it falls onto the telluric absorption 
features of the A- and B-band.
Since the FDF galaxies have varying redshifts different lines are affected
in each case. 
Therefore, the quality of each measurement was checked individually and we 
give quality flags along with the line strengths and their (statistical) 
errors in the Appendix.
The presented values are still not directly comparable to the central values
of local galaxies that were measured on the Lick system.
The reason for this is that the distant galaxies are apparently so small that
the slitlet of width 1\arcsec\ covered a large fraction of the light 
distribution, typically 1--2 half-light radii $R_e$.
The observed line strengths are, therefore, only luminosity-weighted average
values, which are not equivalent to the central values if a radial gradient
across the galaxy exists.
To take this effect into account, the measured values were corrected based on
the logarithmic gradients as they were determined by \citet{MTSBW03}
with high-$S/N$ spectra of Coma cluster galaxies (see also \citet{JFK95}).
The aperture of the 13 FDF galaxies were determined individually considering
both the slitwidth and the number of rows that were averaged by the Horne
extraction algorithm.
The same correction procedure was applied to the velocity dispersions, too.

Structural parameters as well as the total brightness of a galaxy were 
measured on our HST/ACS images of the FDF.
With a mosaic of four pointings, almost the entire field of the FDF was imaged
through the F814W filter by the ACS/WFC with a total exposure time of 2360\,s
per quadrant.
Analysis was performed on the pipeline-reduced images with an additional
cosmic ray rejection procedure.
The two-dimensional surface brightness distribution of a galaxy was fitted 
using the GALFIT package of \citet{PHIR02} to determine total light and 
half-light radius.
Four different fitting functions (pure de Vaucouleurs, pure Sersic with 
variable exponent ($n=1-4$), and these two profiles in combination with an
exponential disk component) were applied.
Comparing the residual images and reduced $\chi^2$-quality values, the best
fit to the observed light distribution of a galaxy was assessed, 
which was in most cases the combination of a Sersic plus disk component.
To test the robustness of the profile analysis, 2D--surface brightness fits 
were also performed using the GIM2D package of \citet{SWVSP02}.
Total magnitudes determined with the two methods agree well within the errors
as well as with those values derived with the SExtractor package \citep{BA96} 
applied on the groundbased $I$ band images.
The magnitudes were calibrated onto the Vega system with a zeropoint of
25.478 and a color term of $0.042 (V-I) + 0.012 (V-I)^2$ 
(M. Sirianni, ESA STScI, priv. com.).
Aperture colors (2\arcsec diameter) were taken from our groundbased 
photometry.
K-corrections were calculated by convolving respective filter transmission
curves with the redshifted SED of the elliptical template galaxy from 
\cite{KCBMS96}.
We checked for internal consistency by transforming observed $F814W$-magnitudes
into rest-frame Johnson-$B$ as well as observed F{\sc ors} filter magnitudes 
that best match the redshifted $B$-band ($g$ for $0.2<z<0.3$ and $R$ for 
$0.3<z<0.7$).
The derived $B_{\rm rst}$ values agree well with each other (differences are
smaller than errors).

Structural parameters, total magnitudes as well as the distance luminosities
for the cosmology we use are given in the Appendix.

\subsection{Elliptical galaxies in distant clusters}

To explore possible differences in galaxy evolution due to
environment, we investigate here in addition to the field galaxies
also early--type galaxies in clusters.  Since the majority of the FDF
galaxies have redshifts around $z\approx0.4$, we take
galaxies from three clusters at $z\approx0.37$: Abell\,370,
CL\,0949$+$4409, MS\,1512.4$+$3647.  The image reduction and data
analysis was already published in \citet{ZB97,BSZBBGH97b,SMGBZ00}.  In
contrast to \citet{ZB97}, we here aperture correct line indices and
the velocity dispersions using the logarithmic gradients to be fully
consistent with the field ellipticals.  All other corrections and the
derivation of the Lick line indices were performed as described above.
Half-light radii and total luminosities were determined on HST/WFPC2
images using the method described by \citet{SBBBCDMW97}.  Combining de
Vaucouleurs and exponential surface brightness profile fits with their
procedure is equivalent to the GALFIT measurements used for the FDF
galaxies, so that the field and cluster samples can be directly
compared to each other here.  Absolute magnitudes and pysical sizes
were re-calculated for the cosmology adopted here.

\section{\label{KSR}The kinematic scaling relations}

\subsection{\label{FJR}The Faber--Jackson relation}

\begin{figure*}
\psfig{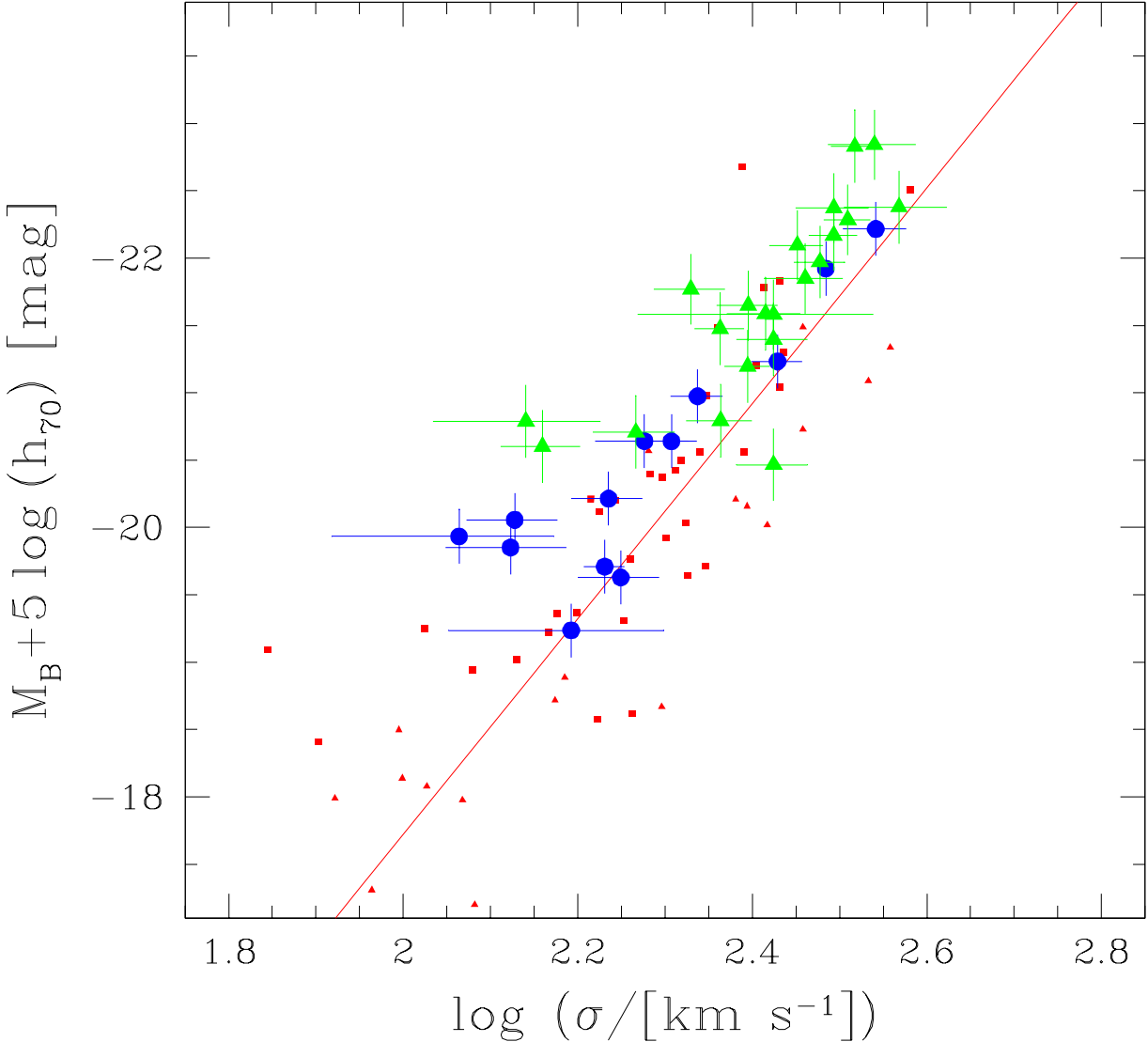}
\caption{\label{fjr}
Faber--Jackson relation between the central velocity dispersion and 
the absolute restframe $B$ magnitude. Local early--type galaxies in the 
Virgo (small
triangles) and Coma (small squares) clusters are from \citet{DLBDFTW87},
the straight line indicates their principal components fit \citep{ZB97}.
Both, ellipticals in distant clusters (large triangles) and in the field (large
circles) follow a similar distribution, which is offset to larger luminosities.
}
\end{figure*}

As the first of the scaling relations that combine the kinematics
(representing both luminous and dark mass) to the stellar populations
(baryons only) of galaxies we investigate the Faber--Jackson relation
\citep[FJR,][]{FJ76}.  Here, the absolute $B$ magnitude and the
velocity dispersion $\sigma$ of local cluster galaxies follow a tight
relation.  In Figure\,\ref{fjr}, we compare the distant galaxies to
the local sample of ellipticals in the Virgo and Coma clusters from
\citet{DLBDFTW87}.  The straight line is a principal components fit to
the nearby galaxies \citep{ZB97}.  For this comparison, we use the
luminosities of the distant galaxies as derived from the ground-based
photometry in order to have a somewhat larger sample (trends are the
same when restricting to space-based magnitudes).  The FDF data points
match the distribution of the distant cluster galaxies very well.
Both distant samples are tight and clearly offset to larger
luminosities with respect to the local FJR.  At low $\sigma$ a few
galaxies with rather large offsets do exist in both samples.
Excluding these outliers, the 17 distant cluster galaxies are brighter
on average for given $\sigma$ by $\langle \Delta M_B^c\rangle = -0.55$
with a standard deviation of 0.25.  This general brightening is
consistent with expectations from passive evolution models if a formation 
redshift of $z_f=4$ is assumed.
For a simple stellar population the predicted $B$-band evolution of such an
object between $z=0$ (age 12\,Gyr) and $z=0.4$ (age 8\,Gyr) 
is $\Delta M_B^m = -0.50$
(using models by \citet{Maras05} for two times solar metallicity
and Salpeter initial mass function).
A similar offset is displayed by the ten FDF ellipticals:
$\langle \Delta M_B^f\rangle = -0.37$ with standard deviation of 0.27.

\subsection{\label{FP}The Fundamental Plane relation}

\begin{figure*}
\psfig{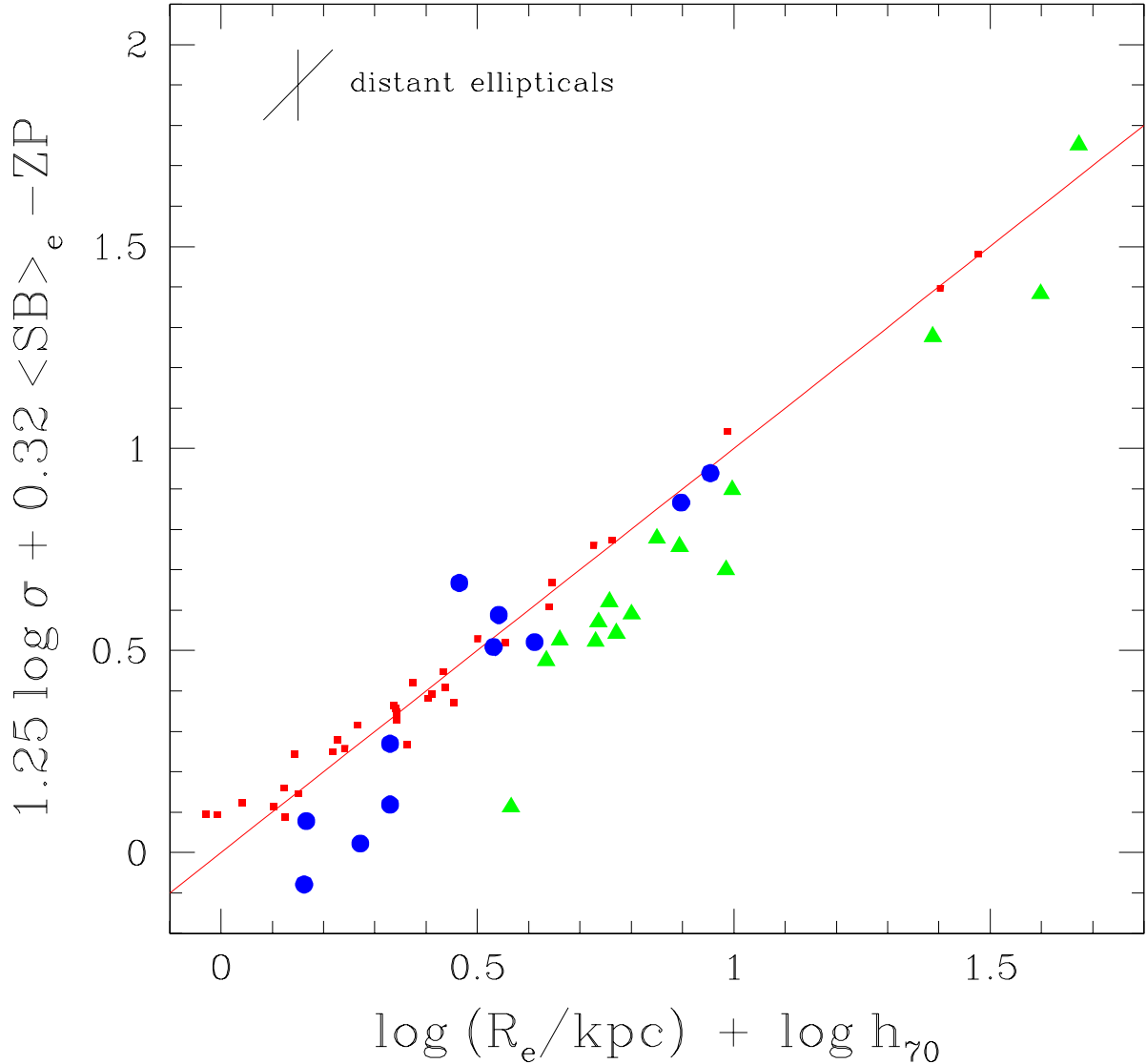}
\caption{\label{fpb} The Fundamental Plane in $B$ seen edge-on.  The
local relation is a principal components fit to
elliptical and S0 galaxies
in Coma.  Ellipticals in distant clusters are indicated by large
triangles, distant field ellipticals by large circles.  A typical
errorbar is displayed in the upper left corner.  }
\end{figure*}

In Figure\,\ref{fpb}, we present the Fundamental Plane (FP) in the
restframe $B$ band.  The edge-on projection is chosen in a way that
the distant-dependent parameter $R_e$ (effective or half-light radius)
is plotted on the x-axis, whereas the distant-independent parameters
$\sigma$ and $\langle SB\rangle _e$ (effective surface brightness) are
displayed along the y-axis.  The local sample which defines the FP
comprises both elliptical and S0 galaxies in the Coma cluster from
\citet{DLBDFTW87}.  Total luminosities and half-light radii of these
galaxies were determined by \citet{SBD93} in exactly the same manner
as our distant cluster galaxies by a simultaneous de Vaucouleurs and
exponential fit to the respective surface brightness profile.  The
straight line indicates the principal components fit from
\citet{BSZBBGH97b}.

The distant galaxy samples displayed here are reduced by two (FDF) and
six (clusters) objects with respect to the FJR analysis, because these
galaxies are not visible on the respective HST images.  Not counting
the galaxy with the smallest and largest radius, the distant cluster
members are on average brighter by $\langle \Delta M_B^c\rangle =
-0.50$ with standard deviation of 0.19.  This assumes that the slope
of the FP does not change within the $\sim4$\,Gyr lookback-time.  This
is equivalent with assuming that the structural parameters $R_e$ and
$\sigma$ do not change within that time and that only the luminosity
of the stellar populations evolves.  The derived value of the global
luminosity increase is then again compatible with a pure passive
evolution of the bulk of the stars that have already old ages
($\sim8$\,Gyr) at the time of observations.

The sample of distant field galaxies lacks the large ellipticals and
cD galaxies present in clusters of galaxies.  On the contrary, the
field sample encompasses some smaller galaxies.  These objects display
the largest luminosity offsets from the local FP, but lie still within
$2\,\sigma$ from the mean offset of the cluster galaxies.  Excluding
the two galaxies with ``positive'' evolution, the average brightening
of the remaining nine FDF galaxies is $\langle \Delta M_B^f\rangle =
-0.34$ with standard deviation of 0.28.

The one outlying galaxy (object FDF\,6336) has a very elongated and
distorted appearance on the HST image.  It's very low Sersic index may
indicate the presence of a disk but the disk/bulge ratio is very
small.  In the Kormendy diagram (relating $R_e$ to $SB_e$) it is also
an outlier with either too faint a surface brightness or too small a
half-light radius.  The same behaviour is exhibited by object
FDF\,7796, which may have an overestimated $\sigma$.  The combined
effect makes this galaxy look ``normal'' in the FP.  But its
appearance is again very elongated and distorted on the ACS image and
it may be associated with a companion galaxy.  The galaxies which are
likely to be lenticular galaxies judging from their appearance do
follow the same distribution in the FP as the ellipticals without any
prominent disk.

\section{\label{MSR}The \mgb--$\sigma$ relation}

\begin{figure*}
\psfig{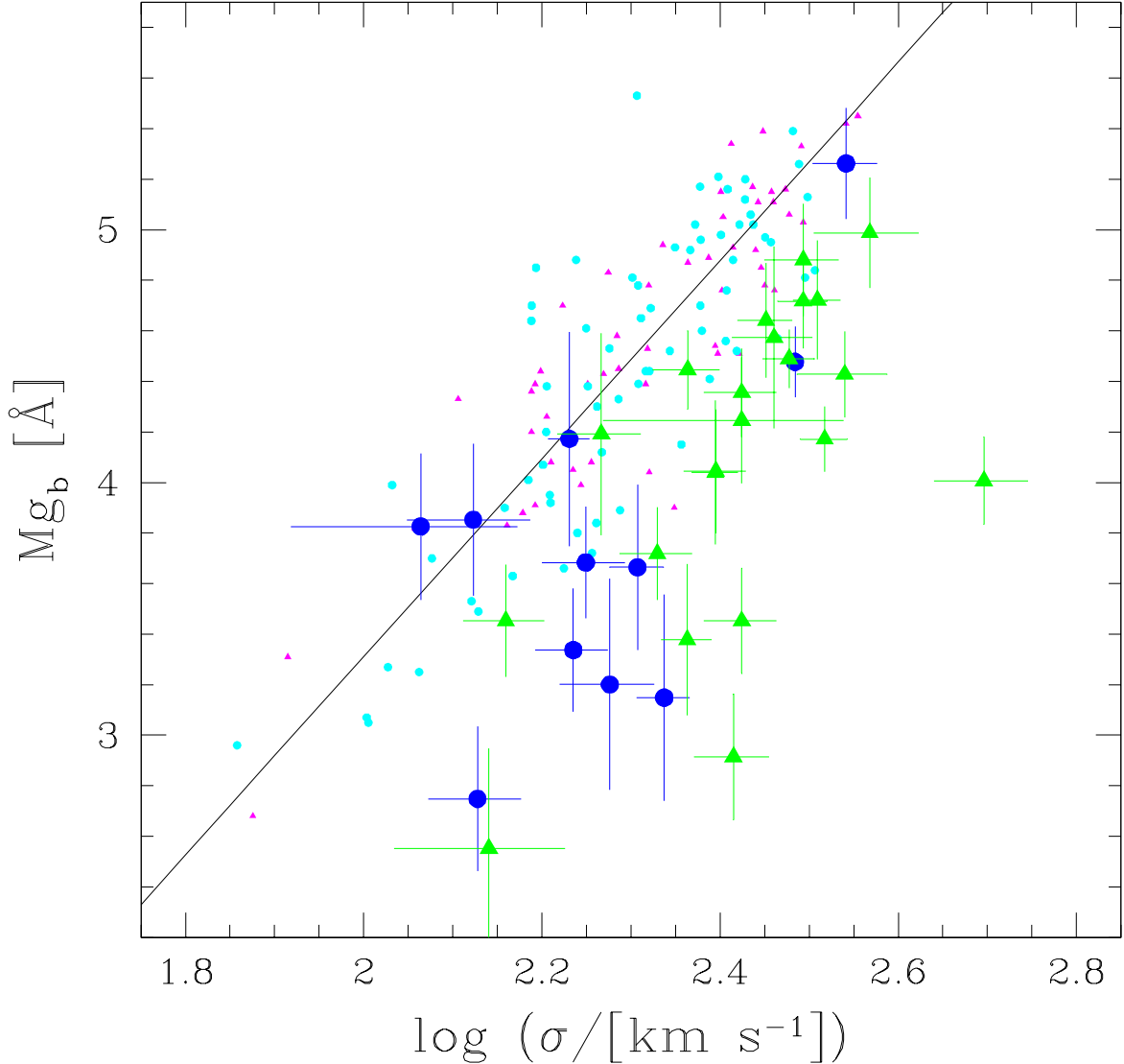}
\caption{\label{msr} \mgb--$\sigma$ relation between the central
velocity dispersion and the central \mgb\ line strength.  Local
galaxies (small symbols) are drawn from the compilation of
\citet{TMBM04} with E and S0 galaxies from
the field (cyan circles) and clusters (magenta triangles).  Both,
ellipticals in distant clusters (green large triangles) and in the
field (blue large circles) follow a similar distribution that is
offset from the local relation (solid line) towards lower
\mgb-strengths for given $\sigma$.  }
\end{figure*}

Another scaling relation that combines the kinematics to the stellar
population is the \mgb--$\sigma$ relation \citep{BBF93}.
In the following sections and figures we compare the distant galaxies
to the recent compilation of 124 early--type local galaxies by
\citet{TMBM04} that contains both morphological classes E and S0 and
encompasses different environments from the field to clusters.  They
combined galaxies with high $S/N$ measurements of absorption line
strengths and velocity dispersions selected from
\citet{Gonza93,MTSBW03,BBMTM02} and \citet{MZTMB05}.  This compilation
has the advantage for us that very similar methods to our procedures
were used for the derivation of indices and $\sigma$ utilizing the
same correction functions.  Furthermore, model parameters of the
stellar populations (see Section\,\ref{sp}) are based on the same
code.  Thus, possible systematic biases between the distant and local
samples are greatly reduced.

In Figure\,\ref{msr}, we show the \mgb--$\sigma$ relation of the local
sample with the distant galaxies overplotted.  We linearily fit the
total local sample by a bisector, since differences between the
various local subpopulations are significantly smaller than the
average offset displayed by both distant samples.  For fixed $\sigma$,
\mgb\ values are smaller with a mean offset of --0.8\,\AA\ in both
cases.  This can be interpreted as evolution in age and metallicity
\citep{BZB96}.  To further explore this, we will compare below the
measured equivalent widths of absorption features to evolutionary
models of stellar populations in so-called line diagnostic diagrams
(Section\,\ref{ldd}).  In Section\,\ref{sp}, we derive model ages and
metallicities and explore their dependence on mass $\sigma$.

For galaxies with $\sigma>250$\kms, the relation of the distant
samples is as tight as for the local comparison galaxies.  Smaller
galaxies display a larger scatter which is compatible with the result
by \citet{TMBM04} that low-mass galaxies have more extended star
formation histories.  
\citet{ZB97} speculated whether some of the cluster galaxies with low
\mgb\ for their $\sigma$ might be post-starburst E$+$A galaxies, where
a recent ($\sim1.5$\,Gyr ago) starburst diluted the \mgb\ feature.
Those E$+$A galaxies are thought to be intermediate between spirals
and ellipticals in the sense that two spirals are transformed into an
early--type galaxy through merging.  Although some of local
representatives of this galaxy type have still discernible disks,
their surface brightness profiles are already bulge dominated
\citep{YZZLM04} and their kinematics pressure supported
\citep{NGZZ01}.  One of the characteristics of E$+$A galaxies is their
relatively strong absorption in the higher Balmer lines like \hd\
compared to quiescent ellipticals.  Unfortunately, we can not access
this issue for the cluster galaxies, because our spectra do not cover
this ``blue'' part.  Also, the [\oiii]\,5007 line, if present in
emission often used as indicator for ongoing star formation or for
correcting the \hb\ absorption for possible emission fill-in, falls
into the atmospheric B band at redshifts around 0.4 and can,
therefore, not be used for this investigation.

In the case of the FDF galaxies, we could measure \hd\ for seven
objects.  But only three (FDF\,1161, 4285, 7459) of the four FDF
galaxies (\& FDF\,6439) with seemingly too low an \mgb\ for their
$\sigma$ have \hd\ measurements.  The other four FDF objects with
known \hd\ (FDF\,5011, 6307, 6336, 7796) are distributed in the
\mgb--$\sigma$ plane like the other ones, although one of them
(FDF\,7796) has rather strong \hd.  Only one of the low-\mgb\ objects
(FDF\,7459) has very strong \hd\ absorption which would be in
compliance with a post-starburst model.

\section{\label{ldd}Line diagnostic diagrams}

\begin{figure*}
\psfig{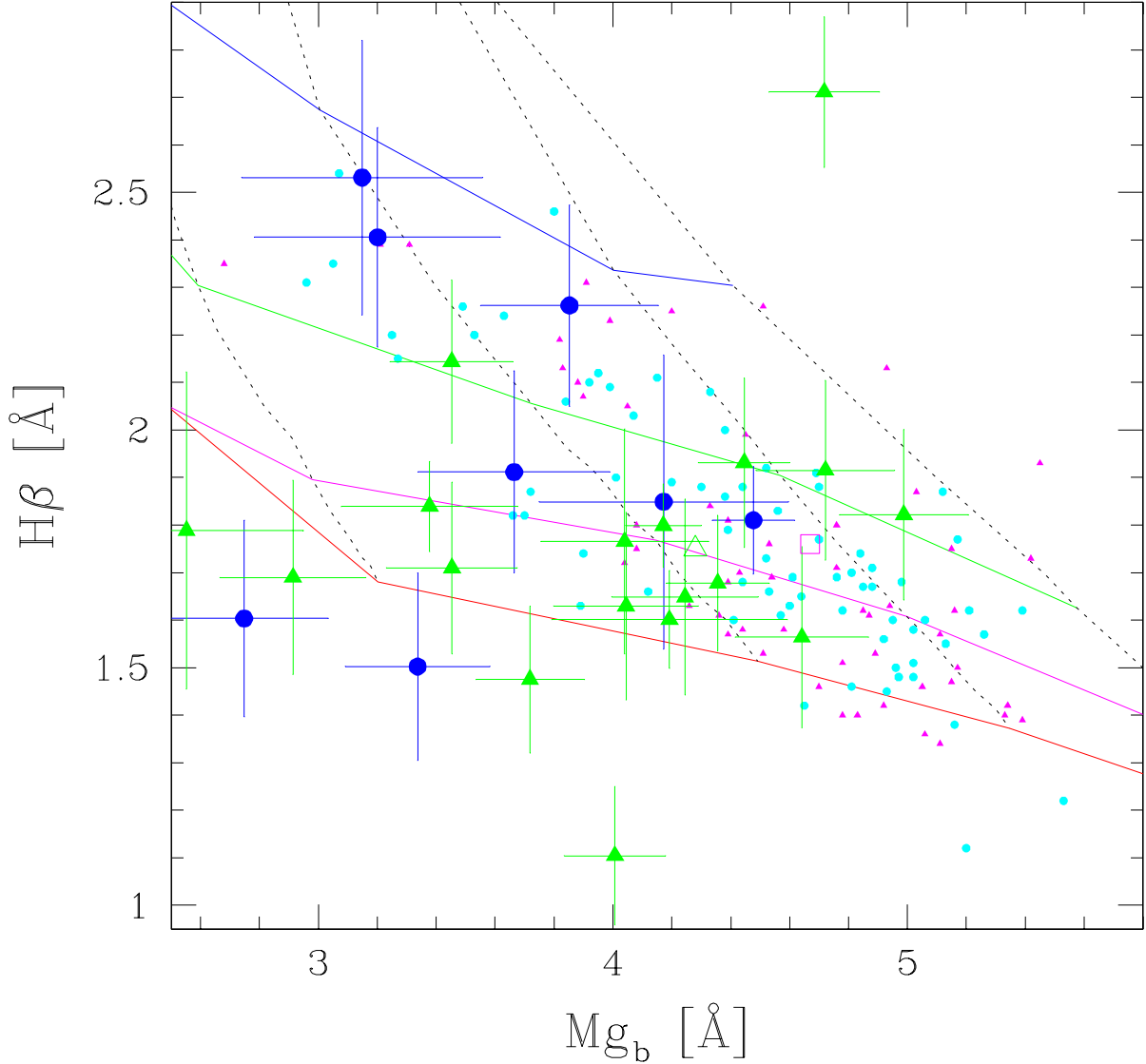}
\caption{\label{mghb}
The same samples (and symbols) as in Figure\,\ref{msr}, displaying 
\hb\ vs. \mgb\ taken
as indicative of age and metallicity [Z/H], respectively.
Big open triangle and square denote the mean value of the distant and local
cluster sample respectively.
The lines connect model values from \cite{TMB03} enhanced in Mg so that
[Mg/Fe]\,$=0.3$. Solid lines correspond to ages of 2, 5, 10, 15\,Gyr top to
bottom and dashed lines to [Z/H]\,$=-0.33, 0, 0.33, 0.67$ left to right.
}
\end{figure*}

\begin{figure*}
\psfig{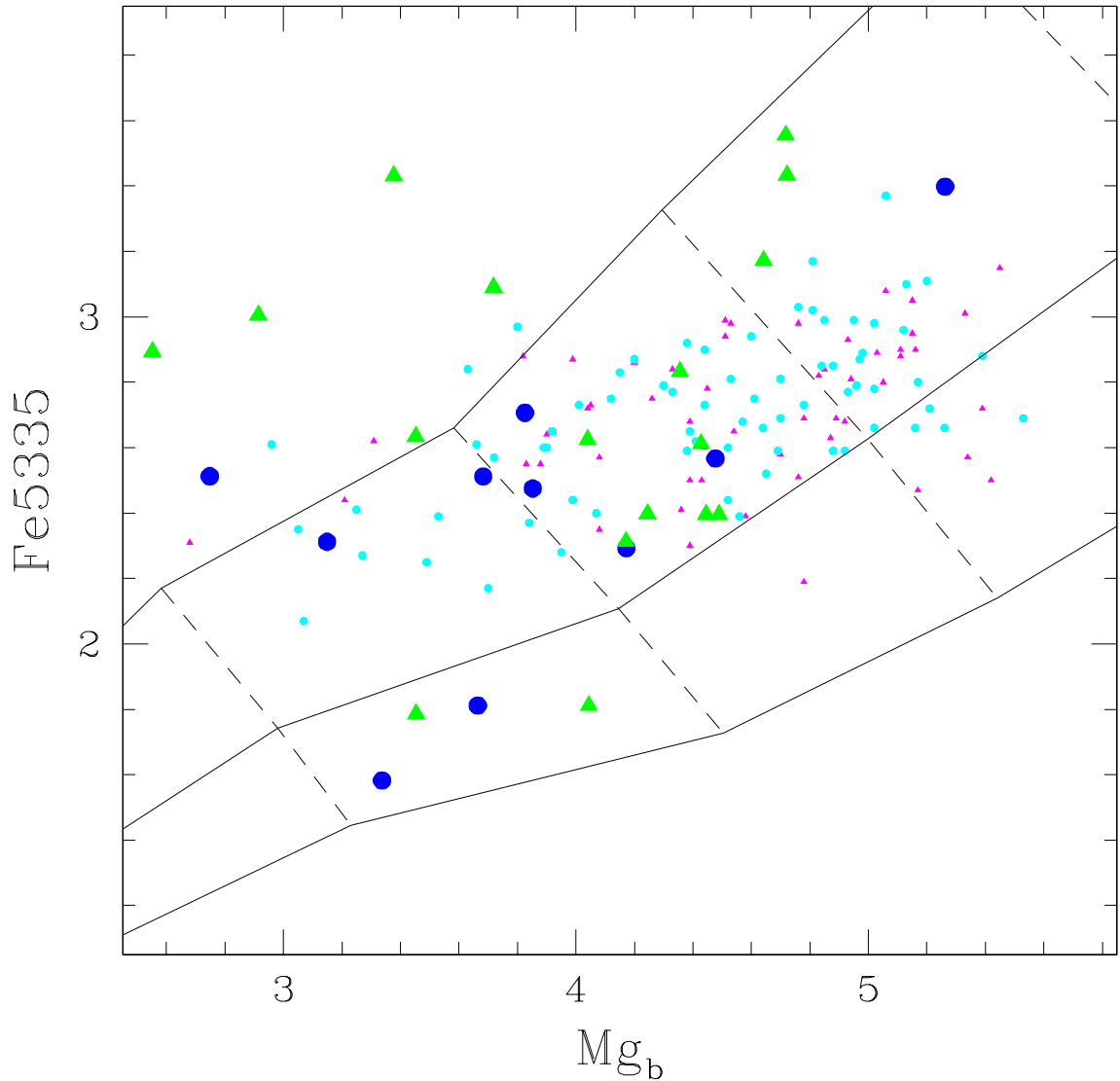}
\caption{\label{mgfe}
The same samples (and symbols) as in Figure\,\ref{msr}, displaying 
\fesix\ vs. \mgb\ to investigate the relative enrichment of Fe and Mg.
The lines connect model values from \cite{TMB03} for a constant age of 10\,Gyr.
Solid lines correspond to [Mg/Fe]\,$=0, 0.3, 0.5$ top to
bottom and dashed lines to [Z/H]\,$=-0.33, 0, 0.33, 0.67$ left to right.
}
\end{figure*}

Absorption line strengths can be taken as measurements of the average
age and metallicity of a stellar population.  Within the Lick/IDS
system, \hb\ usually is taken as an indicator for age, whereas a
combination of \mgb, \fefiv, and \fesix\ yields information on the
mean metallicity [Z/H] \citep[e.g.][]{FFBG85,TWFBG98}.  Since a
galaxy's spectrum is averaged over the slit aperture, any possible
radial gradient of the lines is smeared out and the measurements
represent luminosity-weighted mean values only.  The first effect is
counterbalanced by correcting the observed line strength (see
Section\,\ref{obs}) according to aperture size assuming the validity
of the locally determined radial gradients at higher redshifts.  The
age and metallicity of a stellar population can be deduced by
comparing different indices in so-called line diagnostic diagrams to a
model grid.  Minimizing the differences between observed and model
values, best fitting values for age and metallicity can be determined.
Before we investigate the fitted model parameters in
Section\,\ref{sp}, we show the actual data in Figures\,\ref{mghb} and
\ref{mgfe}.  As mentioned before, important absorption lines of a
galaxy can get redshifted into the regime of strong terrestrial
emission and absorption features, so that it is not possible to
measure them with sufficient accuracy.  For our distant galaxies at
$z\approx0.4$ one or more of the four traditional indices are
affected.  In order not to greatly reduce the samples, we do neither
combine as usual \fefiv\ and \fesix\ to a mean Fe index
$\langle\mbox{Fe}\rangle$ nor do we add \mgb\ to form [MgFe].

To investigate qualitatively age and metallicity effects, we plot in
Figure\,\ref{mghb} only \hb\ versus \mgb\ with a model grid from \citet{TMB03}
that takes into 
account the enhancement of Mg over Fe observed in most elliptical galaxies
by fixing [Mg/Fe] to 0.3.
The local cluster galaxies are mainly located in a region between solar and
thrice solar [Z/H] and old ages of 5--15\,Gyrs \citep{TMBM04}.
The error-weighted mean values for the local \textit{cluster} galaxies are
1.76\,\AA\ for \hb\ and 4.67\,\AA\ for \mgb.
The distant cluster ellipticals show a wide spread in \mgb\ with accordingly 
sub- to supersolar metallicities.
The error-weighted mean values are 1.74\,\AA\ for \hb\  and 4.28\,\AA\ for 
\mgb.
The statistics of the distant field ellipticals suffers from the low number
of eight galaxies.
Three galaxies have intermediate
populations similar to the mean of the distant cluster sample.
Two FDF galaxies have such low \hb\ that they fall below the ``oldest'' model
line (15\,Gyr) and seem to have subsolar [Z/H],
whereas three other ones are compatible with ages of $2-3$\,Gyrs 
with $1-2\times$ solar [Z/H] assuming simple stellar populations (SSPs).

To look for any possible enhancement of Mg over Fe with respect to the solar
abundance ratio [Mg/Fe]\,$=0$, we plot in Figure\,\ref{mgfe} the \fesix\
index versus \mgb.
Local galaxies are located between $1-3\times$ solar [Mg/Fe].
The distant objects scatter within the same range independent from being in
a cluster or field environment.

\section{\label{sp}The stellar populations}

We now discuss the model parameters for the galaxies.  For each
individual galaxy best fitting model parameters were determined by a
minimization algorithm as explained in detail in \citet{TMBM04}.  The
measured ($\sigma$ and aperture corrected) line strengths were
compared iteratively to predicted model line strengths where the
[Mg/Fe] ratio, metallicity [Z/H], and age was varied.  Observed
strengths of \mgb\ and \fesix\ were used in combination whenever their
measurement was trustworthy.  These lines were simultanously fitted
together with one of the Balmer indices \hb, \hgf, or \hdf\ (when
available).  With that procedure we took into account the dependence
of these lines on the Mg/Fe ratio, which is in particular important
for the higher order Balmer lines \citep{TMK04}.

\begin{figure*}
\psfig{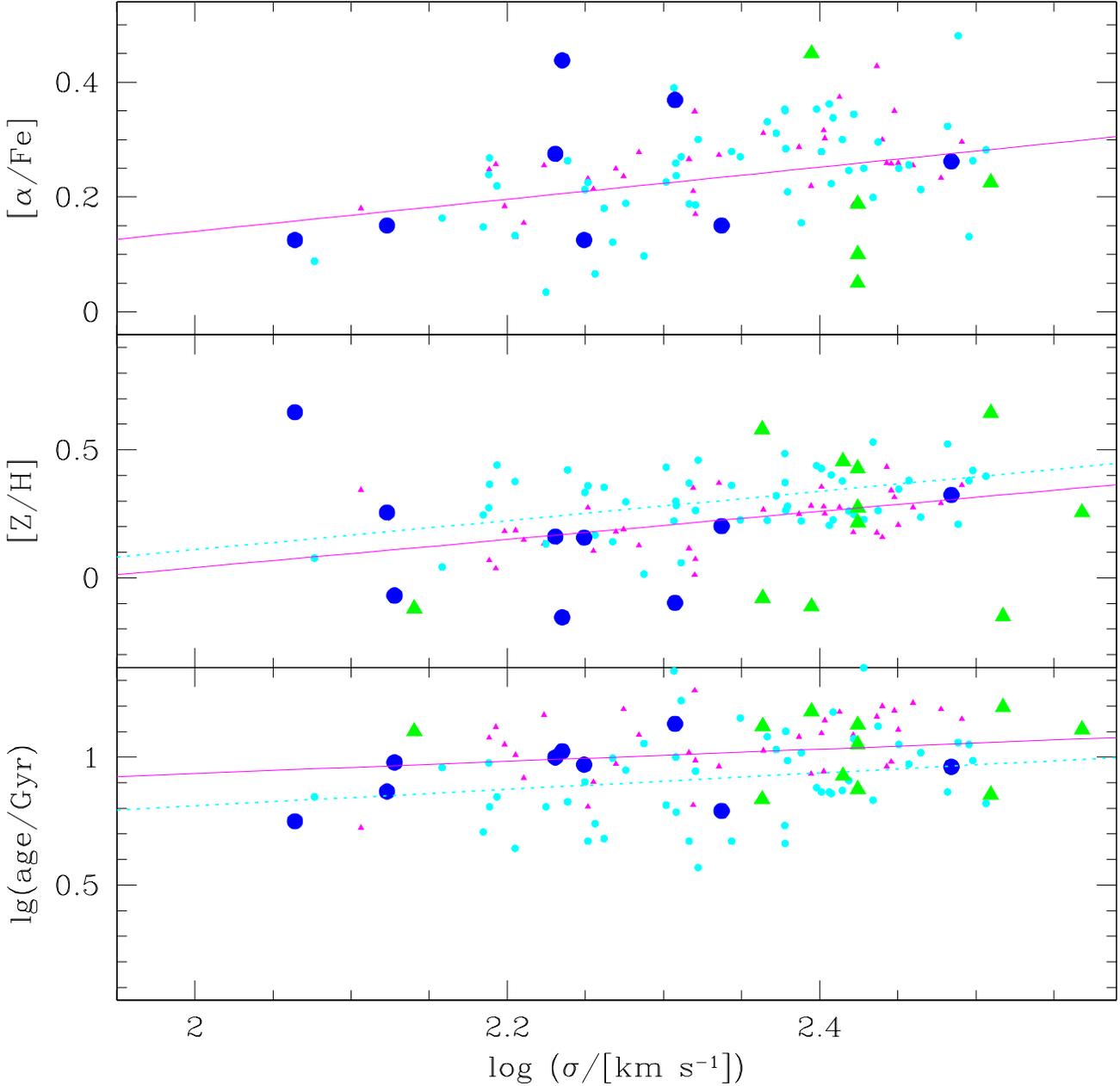}
\caption{\label{sigmod} The samples (and symbols) as in Figure\,\ref{msr}
except that the local sample is now restricted to Category\,1 (ages older
than $\sim5$\,Gyr) galaxies.
The lines indicate the local relations between
age, metallicity [Z/H], and [$\alpha$/Fe] with velocity dispersion as
derived through Monte Carlo simulations by \citet{TMBM04}, separately
for the cluster (solid line) and field sample (dotted line).  
For the distant galaxies, the
respective look-back times have been added to the derived ages.  }
\end{figure*}

In Figure\,\ref{sigmod}, we investigate the scaling relations of the 
resulting stellar parameters with velocity dispersion found for local
early-type galaxies \citep{TMBM04}.  Ages, metallicities, and
$\alpha$/Fe ratios correlate with $\sigma$ and therefore stellar mass.
For the comparison, we now only take the 86 galaxies from the local sample
classififed as being old ($\gtrsim5$\,Gyr),
since the presented scaling relations were
derived for this object class (Category\,1) only.
The local cluster and field samples differ slightly in the zeropoint
of the linear fits for ages and for metallicities.  In
Figure\,\ref{sigmod}, we overplot the derived stellar population
parameters of the distant galaxies.  For this comparison, we
\textit{add the respective look-back time to the age of a 
galaxy}.
For all three distributions, the distant samples nicely match the
local ones.  The individual data points scatter around the local fit
lines, but not significantly larger than the local ones, and follow
the same trends.  In particular, the local $\alpha$/Fe--$\sigma$
relation is also obeyed by the distant galaxies.  This is in
compliance with the derived ages that are younger only corresponding
to their look-back times.  From this we can conclude that for the
majority of the local galaxies no significant star formation episodes
and no further chemical enrichment has taken place within the
last $\sim5$\,Gyr.  Slight differences (which are not significant) in
[Z/H]--$\sigma$ between the local and distant samples may be
attributed to the fact, that the aperture correction in the case of
the distant galaxies was done onto the nominal aperture whereas the
local galaxies were measured within 1/10th of $R_e$.  Since age and
$\alpha$/Fe have zero radial gradients in local galaxies
\citep{MTSBW03}, only our [Z/H] determinations could be affected in
the sense that they are systematically slightly too low.

\section{\label{su}Summary and discussion}

We have investigated the kinematic, photometric, structural, and stellar
population parameters of a sample of field galaxies with mean redshift 0.4.
The galaxies were drawn from the F{\sc ors} Deep Field on the basis of their
early--type SED, i.e. their optical and near-infrared colours.
It turned out that they all have indeed absorption-line spectra without any
emission line, i.e. without obvious indication for ongoing star formation.
Their structure is mostly dominated by a bulge component, but a few galaxies
clearly have disks like local lenticular S0 galaxies.

To learn about any possible evolution specific to the low-density
environment, we compare the FDF sample to galaxies in three rich
clusters at similar redshifts and to local samples both with field and
cluster galaxies.  We have first explored the luminosity evolution via
the Faber--Jackson and Fundamental Plane relations.  Assuming that the
structure (size) and mass (velocity dispersion) of elliptical galaxies
do not change with time, we find consistently from both scaling
relations a very modest evolution in brightness that can be modeled by
passive evolution of a simple stellar population (SSP).  The average
increase of the $B$ luminosities by 0.3--0.5\,mag is found for both
the distant field and cluster galaxies and can be explained by a high
formation redshift ($z_f>2$) of their stars.

This result is consistent with those studies that also find a weak
luminosity evolution and little difference between field and cluster
ellipticals.  In their FP study of 18 field early--type galaxies with
$\langle z\rangle=0.4$, \citet{DFKI01} also find that they are
brighter in $B$ by 0.4\,mag in general, which can be modeled with an
SSP only slightly younger than in the case of their cluster galaxies.
A similar conclusion was reached by \citet{RKFKM03}, who measured an
increase by 0.5\,mag at $z=0.4$.  In contrast, other groups like
\citet{TSCMB02} or \citet{GFKIS03} find that the evolution for field
galaxies is stronger compatible with a significant younger mean age
for their stellar populations than in the case of cluster galaxies.
Restricting their samples to galaxies around $z=0.4$, $\Delta B$
amounts to 0.7--0.8\,mag.  The differences among the various studies
become even more prominent at higher $z$, where the latter authors
derive a much larger brightening for the field galaxies in comparison
to cluster members.  The rather small differences at $z\approx0.4$ may
arise from the still quite small numbers of observed galaxies per
sample.  These small samples may therefore also be more subject to the
particular method to select the targets.  Our FDF galaxies are, for
example, redder by 0.7\,mag on average in observed $(V-I)$ color than
the Gebhardt et al. sample, which has $\langle (V-I)\rangle=1.3$ for 5
ellipticals at $0.3<z<0.5$, while both our field and cluster galaxies
have $\langle (V-I)\rangle=2.0$.
This indicates that our field galaxies probably trace predominantly the 
upper envelope of the age--$\sigma$ distribution.

Compatibility with SSP models of passive evolution is also found for
the majority of the FDF galaxies in the \mgb--$\sigma$ relation.
Their mean offset from the local relation is with $-0.8$\,\AA\
identical to the one for the distant cluster galaxies.  
The local slope is followed by the distant galaxies as well
with a tendency for a slight steepening at the low-mass end.
This indicates that age cannot be the
principal driver of the main part of the relationship. Otherwise a
significant steepening also at the high-mass end would be detected, as
the \mgb\ index of younger objects,
that also have on average lower $\sigma$ according to the age--$\sigma$
correlation, would evolve
faster to lower index values with increasing redshift. This result is
in line with the conclusion drawn by several local studies
\citep[e.g.][]{TFWG00a,KSCDK02,TMBM04} 
that the \mgb--$\sigma$ relation is driven by metallicity rather than by age.

To explore in
more detail the distribution in metallicity, chemical enrichment, and
mean ages of the stellar systems, we compare diagnostic absorption
lines from the Lick system to SSP models.  To this purpose we
simultaneously fit measured Mg, Fe, and Balmer indices by varying
model ages, [Z/H] metallicities, and [$\alpha$/Fe] ratios.  We find a
broad range in these model parameters for both distant samples, field
and clusters.  But their distributions are not random because all
three stellar population parameters scale with velocity dispersion,
and, therefore, to some degree with total mass, like their local
counterparts.

A passive evolution of cluster galaxies from $z=0.5$ to 0 was also
favoured by \citet{JSC00} investigating the distribution of \hda\ with
metal lines measured in co-added spectra.  They detect no significant
difference between morphologically classified S0 and E galaxies
arguing also, that the majority of S0 cluster members do not show any
evidence of recent star formation activity as expected in some
scenarios of galaxy transformation.  Two of our FDF galaxies
(FDF\,6336 \& 7796) have extraordinary \hdf\ absorption leading to
young model ages under the assumption of a single stellar population.
Their strengths may also be explained in a scenario where these
galaxies had experienced a short intense star burst in their recent
past involving only a small fraction of their mass.  The ACS images
reveal a significant disk component in these two galaxies, which may
point to a very early spiral morphology.  In that case a low-level
continous star formation might be expected, but we do not detect any
emission line.

\citet{KIFD01} investigated the higher-order Balmer lines \hda\ and
\hga\ of early--type galaxies in three clusters at $z=0.3,0.6$ \& 0.8
claiming that there is a correlation with $\sigma$ with small scatter.
Interpreting the offsets in the zeropoint for the three redshifts as
caused by an evolution in the mean age of the stellar systems, they
derive formation redshifts of $z_f\gtrsim2$ for SSP models assuming
co-evality for all galaxies.  Since we only have four FDF galaxies
with high-quality measurements of the two indices (and none in the
cluster sample), we can not confirm or exclude a correlation with
$\sigma$.  But the assumption that all galaxies have roughly the same
age must be rejected for our cluster galaxies.

\begin{figure*}
\psfig{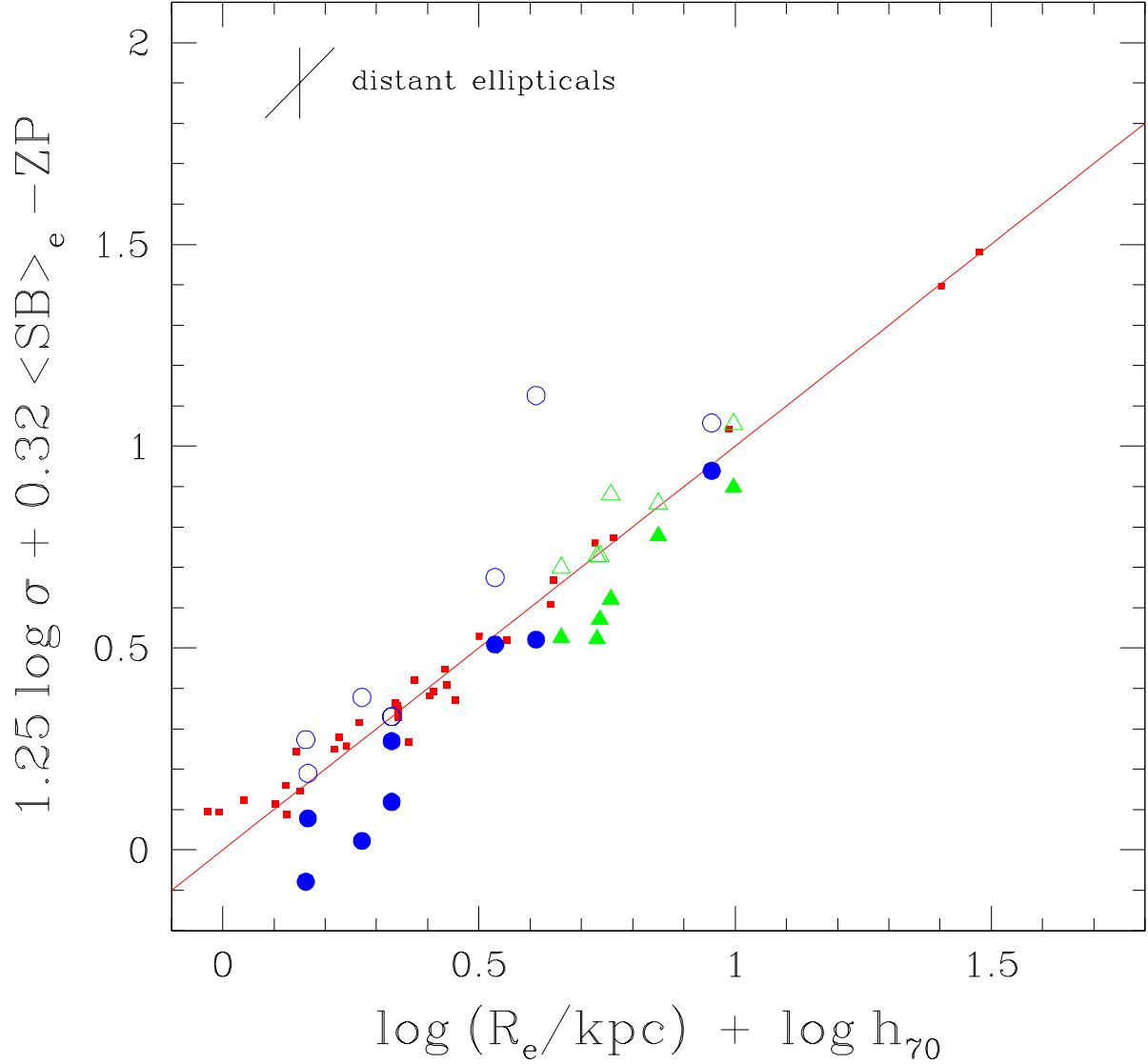}
\caption{\label{fpevo} The Fundamental Plane as in Figure\,\ref{fpb}
but restricted to those galaxies for which we could derive light-averaged
model ages of their stellar population.
The open symbols represent the data ``corrected'' for the predicted
brightening for each galaxy and nicely match the distribution of the local
galaxies.
}
\end{figure*}

At last, we investigate whether our findings derived from the Fundamental 
Plane and the absorption line diagnosis with respect to luminosity evolution
are compatible with each other also quantitatively.
We determine the model brightening of our galaxies from the predicted
restframe $B$ magnitudes for the respective ages and lookback times
taking into account the derived [Z/H] metallicities and [$\alpha$/Fe] ratios
using again SSP models by \citet{Maras05}.
In Figure\,\ref{fpevo}, we make this comparison by subtracting the predicted
luminosity evolution from the observed data points.
The evolution-corrected distant FP falls onto the local relation and nicely
matches the distribution of the local galaxies.
The one outlier is that galaxy (FDF\,7796), for which we speculated above
that it has undergone a recent starburst therby causing the light-averaged
age to be very low.

In summary, we conclude from our study that the evolution of the
stellar populations of the majority of early--type galaxies within the
last 3--6\,Gyr is compatible with the scenario of passive evolution.
Since the $\alpha$/Fe--$\sigma$ and age--$\sigma$ relations did not
change since $z=0.4$, no significant star formation and chemical
enrichment has occured since that time.  To investigate certain issues
in more detail, like any possible difference between low- and
high-density environments, new studies with more objects are required
with high $S/N$ measurements allowing the investigation of several
aspects (FP \& line diagnostics) ideally out to higher redshifts.  To
this respect, we are currently analyzing another sample of 
distant absorption-line galaxies drawn from the William--Herschel Deep
Field for which we also have VLT spectra and HST/ACS imaging with a
similar quality as for the FDF galaxies presented here.


\begin{acknowledgements}
We thank the anonymous referee for a very quick response.
We acknowledge fruitful discussions with all FDF consortium members,
in particular with Dr. A. Gabasch, M. Pannella, Dr. R. P. Saglia, Dr. S. Seitz
(all Munich \& Garching) and Dr. K. J\"ager and B. Gerken (G\"ottingen).
We thank J. Fliri and A. Riffeser (Sternwarte Munich) for applying their
cosmic ray removal algorithm to the HST/ACS images.
We are grateful to Drs. J. Heidt, D. Mehlert, and S. Noll (Landessternwarte
Heidelberg) for performing the FDF spectroscopic observations and thank ESO 
and the Paranal staff and the STScI and ST--ECF staff for efficient support. 
We also thank the PI of the FORS project, Prof. I. Appenzeller (Heidelberg), 
and Prof. K. J. Fricke (G\"ottingen) for providing guaranteed time for this
project.
This work has been supported by the Volkswagen Foundation (I/76\,520)
and the BMBF/DLR (50\,OR\,0301)
and made use of the ADS, astro--ph, and CDS data bases.
\end{acknowledgements}


\bibliographystyle{aa}


\appendix

\section{Parameters of the FORS Deep Field galaxies}

\begin{table*}
  \caption[]{Basic parameters for the FORS Deep Field galaxies.\newline
  1. ID from \citet{HAGJS03}, 2. redshift, 3. luminosity distance, 4. 
  lookback time,
  5. signal/noise, 
  6. velocity dispersion corrected for aperture (see Section\,\ref{obs}), 
  7. its error, 
  8.--11. FORS total magnitudes (SExtractor's \textsc{mag\_best}),
  12. absolute restframe B magnitude, 
  \hfill 3., 4., \& 12. for $\Omega_{\rm matter}=0.3$, $\Omega_{\lambda}=0.7$,
  $H_0 = 70$\,km\,s$^{-1}$\,Mpc$^{-1}$\newline
  *: no entry because object not visible on respective image
  }
  \label{hst}
     $$
  \begin{tabular}{lcccrcrccccc}
  \hline
  \noalign{\smallskip}
FDF & $z$ & $dl$ & $t_{lb}$ & $S/N$ & $\sigma$ & $\Delta\sigma$ & 
 $B_{tot}$ & $g_{tot}$ & $R_{tot}$ & $I_{tot}$ & $M_B$ \\
 & & & [Gyr] & & [km/s] & [km/s] & [mag] & [mag] & [mag] & [mag] & [mag]  \\
  \noalign{\smallskip}
  \hline
  \noalign{\smallskip}
1161 & 0.40 & 2147 & 4.2 &  19 & 134 & 16 & 23.4 & 22.3 & 20.6 & 19.9 & -20.2 \\
4285 & 0.40 & 2145 & 4.2 &  29 & 172 & 16 & 23.2 & 22.1 & 20.4 & 19.6 & -20.4 \\
5011 & 0.65 & 3922 & 6.0 &  15 & 268 & 18 & 23.9 & 23.0 & 21.5 & 20.9 & -21.0 \\
5908 & 0.22 & 1112 & 2.7 & 127 & 305 &  5 & 19.7 & 18.4 & 17.1 & 16.5 & -21.9 \\
6307 & 0.45 & 2508 & 4.7 &  29 & 203 & 14 & 23.5 & 22.3 & 20.5 & 19.7 & -20.7 \\
6336 & 0.46 & 2561 & 4.7 &   8 & 156 & 43 & 24.7 & 23.7 & 21.9 & 21.1 & -19.4 \\
6338 & 0.41 & 2238 & 4.4 &  22 & 178 & 19 & 23.9 &  *   &  *   & 20.3 & -19.7 \\
6439 & 0.40 & 2145 & 4.2 &  41 & 217 & 15 & 22.4 & 21.4 & 19.7 & 18.9 & -21.0 \\
7116 & 0.46 & 2544 & 4.7 &  28 & 133 & 21 & 24.2 & 23.0 & 21.3 & 20.5 & -20.0 \\
7459 & 0.54 & 3100 & 5.3 &  20 & 189 & 23 & 24.3 & 22.9 & 21.2 & 20.3 & -20.6 \\
7796 & 0.41 & 2231 & 4.4 &  15 & 116 & 33 & 23.6 & 22.5 & 20.8 & 20.1 & -20.1 \\
8372 & 0.23 & 1142 & 2.7 &  57 & 170 &  9 & 21.6 & 20.7 &  *   & 18.5 & -20.1 \\
8626 & 0.41 & 2226 & 4.4 &  27 & 348 & 29 & 21.4 &  *   &  *   & 17.7 & -22.4 \\
  \noalign{\smallskip}
  \hline 
  \end{tabular}
     $$
\end{table*}
%


%
\begin{table*}
  \caption[]{Line strengths of fully corrected (see Section\,\ref{obs})
    Lick indices of the FORS Deep Field galaxies
    together with their errors and (by-eye) quality flags. Units are \AA.
    Quality flags: 0: satisfactory, 1: a bit noisy, 2: in region of strong sky
    line, 6: affected by telluric B band, 7: affected by telluric A band,
    8: affected by end of spectrum, 9: not visible.
  }
  \label{lick}
     $$
  \begin{tabular}{ccccccccccccc}
  \hline
  \noalign{\smallskip}
 FDF & \hdf & d\hdf & q\hdf & \hgf & d\hgf & q\hgf & \hb & d\hb & q\hb & \mgb & d\mgb & q\mgb \\
  \noalign{\smallskip}
  \hline
  \noalign{\smallskip}
1161 & 1.3 & 0.2 & 0 & -0.6 & 0.2 & 0 & 1.6 & 0.2 & 0 & 2.7 & 0.3 & 1  \\ 
4285 & 1.2 & 0.2 & 0 & -0.7 & 0.2 & 0 & 1.5 & 0.2 & 0 & 3.3 & 0.2 & 0  \\ 
5011 & 1.0 & 0.2 & 0 &  0.4 & 0.2 & 0 & 1.9 & 0.3 & 1 &  *  &  *  & 9  \\ 
5908 &  *  &  *  & 9 &   *  &  *  & 9 & 1.8 & 0.1 & 0 & 4.5 & 0.1 & 0  \\ 
6307 & 0.8 & 0.2 & 2 & -0.8 & 0.3 & 2 & 1.9 & 0.2 & 0 & 3.7 & 0.3 & 0  \\ 
6336 & 2.6 & 0.3 & 0 & -0.7 & 0.3 & 2 & 0.5 & 0.3 & 2 &  *  &  *  & 7  \\ 
6338 &  *  &  *  & 9 &   *  &  *  & 8 &  *  &  *  & 6 & 3.7 & 0.2 & 0  \\ 
6439 &  *  &  *  & 9 &   *  &  *  & 8 & 2.5 & 0.3 & 0 & 3.1 & 0.4 & 0  \\ 
7116 &  *  &  *  & 9 & -0.4 & 0.2 & 2 & 2.3 & 0.2 & 0 & 3.9 & 0.3 & 1  \\ 
7459 & 2.5 & 0.2 & 2 & -0.1 & 0.2 & 0 & 2.4 & 0.2 & 0 & 3.2 & 0.4 & 1  \\ 
7796 & 2.0 & 0.2 & 0 &  1.0 & 0.2 & 0 &  *  &  *  & 6 & 3.8 & 0.3 & 0  \\ 
8372 &  *  &  *  & 9 &   *  &  *  & 8 & 1.8 & 0.3 & 0 & 4.2 & 0.4 & 2  \\ 
8626 &  *  &  *  & 9 &   *  &  *  & 8 &  *  &  *  & 6 & 5.3 & 0.2 & 0  \\ 
  \noalign{\smallskip}
  \hline 
  \end{tabular}
     $$
\end{table*}
%


%
\begin{table*}
  \caption[]{Fully corrected \fesix\ Lick index and SSP model ages, 
    metallicities [Z/H], and Mg/Fe ratios [$\alpha$/Fe] of the FORS Deep Field
    galaxies. 
    Average errors in $\log$\,(age), [Z/H], and [$\alpha$/Fe] are
    0.18\,dex, 0.2\,dex, and 0.15\,dex, respectively.
    In column\,8 it is indicated, which Balmer line was used
    to derive the model age.
    The last four columns are based on the HST/ACS images:
  ACS total I magnitude (\textsc{galfit}), 
  mean surface brightness in B within $R_e$ corrected 
  for cosmological dimming, and half-light radius (\textsc{galfit}).
  Physical units for $\Omega_{\rm matter}=0.3$, $\Omega_{\lambda}=0.7$,
  $H_0 = 70$\,km\,s$^{-1}$\,Mpc$^{-1}$.
}
  \label{mod}
     $$
  \begin{tabular}{cccccrrlcccc}
  \hline
  \noalign{\smallskip}
 FDF & \fesix & d\fesix & q\fesix & age & [Z/H] & [$\alpha$/Fe] & Balmer
  & $I_{tot}$ & $\langle\mu_B\rangle_e$ & $R_e$ & $\log R_e$\\
     & \AA    & \AA     &         & Gyr &       &               &  line
  & [mag] & [mag/sq.as.] & [arcsec] & [kpc] \\
  \noalign{\smallskip}
  \hline
  \noalign{\smallskip}
1161 & 2.5 & 0.3 & 1 & 5.3 & -0.07 & -0.09&  \hgf  & 19.9 & 20.3 & 0.40 & 0.33 \\ 
4285 & 1.6 & 0.3 & 2 & 6.3 & -0.15 &  0.44&  \hdf  & 19.9 & 20.3 & 0.40 & 0.33 \\ 
5011 &  *  &  *  & 9 &  *  &   *   &   *  &  *     & 20.8 & 20.6 & 0.50 & 0.54 \\ 
5908 & 2.6 & 0.2 & 0 & 6.5 &  0.32 &  0.26&  \hb   & 16.3 & 21.5 & 2.50 & 0.95 \\ 
6307 & 1.8 & 0.5 & 0 & 8.8 & -0.10 &  0.37&  \hb   & 19.8 & 20.8 & 0.59 & 0.53 \\ 
6336 &  *  &  *  & 9 &  *  &   *   &   *  &  *     & 21.2 & 21.8 & 0.50 & 0.46 \\ 
6338 & 2.5 & 0.3 & 0 & 5.0 &  0.16 &  0.12&  \hb   &  *   &  *   & *    & *    \\ 
6439 & 2.3 & 0.5 & 0 & 1.9 &  0.20 &  0.15&  \hb   & 19.1 & 19.2 & 0.35 & 0.27 \\ 
7116 & 2.5 & 0.4 & 2 & 2.6 &  0.25 &  0.15&  \hb   & 20.6 & 19.7 & 0.25 & 0.16 \\ 
7459 &  *  &  *  & 9 &  *  &   *   &   *  &  *     &  *   &  *   & *    & *    \\ 
7796 & 2.7 & 0.4 & 0 & 1.3 &  0.65 &  0.12&  \hgf  & 20.1 & 21.8 & 0.75 & 0.61 \\ 
8372 & 2.3 & 0.4 & 0 & 7.2 &  0.16 &  0.28&  \hb   & 18.6 & 19.8 & 0.40 & 0.17 \\ 
8626 & 3.4 & 0.2 & 1 &  *  &   *   &   *  &  *     & 17.9 & 21.0 & 1.45 & 0.90 \\ 
  \noalign{\smallskip}
  \hline 
  \end{tabular}
     $$
\end{table*}

\end{document}